\documentclass
[%
reprint,
 amsmath,amssymb,floatfix,
 aps,
prb,
floatfix,
]{revtex4-2}

\usepackage{graphicx}
\usepackage{dcolumn}
\usepackage{bm}
\usepackage{physics}
\usepackage{svg}
\usepackage{lipsum}
\usepackage{comment}
\usepackage[normalem]{ulem}
\usepackage{cancel}
\usepackage{hyperref}
\hypersetup{
    colorlinks=true,
    citecolor=blue, 
    linkcolor=blue, 
    filecolor=blue,      
    urlcolor=blue,
    }

\begin{document}

\preprint{APS/123-QED}

\title{Higher-order discrete time crystals and enhanced sensing in a quantum kicked top}

\author{Subhashis Das}
\email{sdresearchworks64@gmail.com}
\affiliation{Department of Physics, Indian Institute of Technology, Hyderabad 502284, India}
\author{Vishal Khan}
\email{vishal.khan.research@gmail.com}
\affiliation{Department of Physics, Indian Institute of Technology, Hyderabad 502284, India}
\author{Atanu Rajak}
\email{atanu@phy.iith.ac.in}
\affiliation{Department of Physics, Indian Institute of Technology, Hyderabad 502284, India}

\date{\today}

\begin{abstract}
We characterize various dynamical phases of the simplest version of the quantum kicked top model, a paradigmatic system for studying quantum chaos, which exhibits both regular and chaotic behavior depending on the kick strength. In a previous study, the existence of higher-order discrete time crystals (DTCs) was observed in an infinite-range interacting $p$-spin model, where it was proposed that the order of the DTC satisfies the relation $q\le p$~\cite{14}. Within this framework, the $p=2$ model is expected to host only a $2$-DTC phase. However, interestingly, we demonstrate here the existence of a robust $4$-DTC phase in the quantum kicked top, which effectively corresponds to a $p=2$ model with infinite-range interactions. We also show that the system hosts robust $2$-DTC and dynamical freezing (DF) phases around alternating rotationally symmetric points. 
We explain the emergence of higher-order DTC phases through the classical phase portraits of the system, connected with spin coherent states (SCSs), by identifying special islands that arise within a specific parametric regime. Unlike the $2$-DTC phase, the $4$-DTC phase appears only for certain initial states, as demonstrated through exact calculations. The robustness of the $4$-DTC phase is further investigated through the dynamics of the linear entropy as a function of the angular momentum.
We also find an emergent conservation law for both the $2$-DTC and DF phases, while no dynamical conservation arises periodically for the $4$-DTC phase. By investigating the quantum Fisher information, we also demonstrate enhanced metrological sensitivity at the boundaries between different dynamical phases for the estimation of system parameters.

\end{abstract}

\maketitle


\section{\label{sec:level1}introduction}
In recent years, there has been considerable interest in studying quantum many-body systems driven out-of-equilibrium, both in theory and in experiments. Specifically, periodically driven quantum systems have attracted much attention to control the physical properties of the system and to provide exotic non-equilibrium phases with no classical counterparts, generically known as Floquet engineering~\cite{bukov2015universal,nag14,oka2019floquet,rudner2020band,harper2020topology}. One of the most interesting Floquet phases is a discrete time crystal (DTC) that is expected to be useful in quantum technological applications~\cite{khemani2016phase,yao2017discrete,else2016floquet,29,khemani2019brief,28,ojeda2021emergent}. A solid state crystal is formed when space translational symmetry is broken. In a similar fashion, a DTC is characterized by breaking the discrete time-translation symmetry followed by subharmonic oscillations of the physical observables. The most fruitful setting for realizing such phases is found to be periodically driven systems where the Hamiltonian has some time periodicity that is broken, manifested by the response of an observable oscillating with integer multiple of time period of the drive. The oscillations persist forever, on approaching the thermodynamic limit. Interactions are essential to get a robust DTC phase in the presence of driving by breaking the time-translation-symmetry of the system~\cite{khemani2019brief,sacha2020time}. However, interacting Floquet systems generically heat up by constant absorption of energy from the drive, thereby facing a serious challenge for stabilizing a DTC phase~\cite{d2014long,lazarides2014,choudhury2014stability}. Therefore, the initial work relies on many-body localizations to realize DTCs to preclude Floquet heating~\cite{zhang2017observation}. In recent years, however, several studies demonstrate DTCs in interacting clean systems without disorder. A few examples include the driven Lipkin-Meshkov-Glick model~\cite{2}, all-to-all interacting spin models with $p$-body interactions~\cite{14} and central spin models~\cite{biswas2025discrete}. The existence of DTC phases has also recently been observed in experiments for both disordered~\cite{zhang2017observation} and cleaned cases~\cite{kyprianidis2021observation}.

Although, in principle, the DTC can have periodicity $nT$ with $n$ being an integer and $T$ is the time-period of the Hamiltonian, the most common case is $n=2$. Usually, this type of DTC is observed in systems of driven interacting spin-$1/2$ particles that have $\mathbb{Z}_2$ symmetry. However, recently, higher-order DTCs have been proposed in bosonic or higher-spin systems that do not have natural $\mathbb{Z}_2$ symmetry~\cite{choi2017observation,yao2017discrete,pizzi2019period}. In addition, higher-order DTCs have been reported in clock models~\cite{surace2019floquet} as well as in spin-$1/2$ systems~\cite{bomantara2021quantum,27}. Recently, it has been shown that higher order DTCs can be found in driven spin systems with infinite-range $p$-body interactions for $p>2$. In the $p$-spin model, it has been argued that the system can host only subharmonic responses of order $q\le p$~\cite{14}. However, in this work, we demonstrate that higher-order DTC phases can exist even for $p=2$. Dynamical Freezing (DF) is another non-equilibrium phenomenon that has been investigated extensively in the last two decades~\cite{23,21,20}. In this case, the systems have some emergent conserved operators whose expectations remain approximately close to initial values with some fluctuations that do not grow with the time. Generally, the emergent conservations appear for strong driving fields which are not present in the undriven system.

In parallel, periodically driven systems provide a promising platform for quantum metrology, as their dynamical phases can be manipulated to achieve enhanced sensitivity in parameter estimation. Quantum metrology has emerged as an indispensable tool in quantum technologies for the high-precision estimation of physical parameters~\cite{taylor2008high,giovannetti2011advances,sidhu2020geometric,montenegro2025quantum}. In this context, the quantum Fisher information (QFI) serves as a fundamental measure for quantifying the precision of a quantum sensor. Recent studies have demonstrated that quantum phase transitions can significantly enhance quantum sensing performance, with the QFI exhibiting pronounced peaks near the critical points~\cite{ilias2022criticality,sarkar2025exponentially,ghosh2025quantum}.

In this work, we address the question of whether one of the simplest quantum chaotic systems can host higher-order and robust discrete time crystal (DTC) phases. We consider the quantum kicked top, a paradigmatic model of quantum chaos~\cite{1,chaudhury2009quantum,krithika2019nmr}. This system possesses a well-defined classical counterpart that exhibits chaotic behavior beyond a certain kicked strength threshold. Below this threshold, the phase space shows a mixed structure where both regular and chaotic classical trajectories coexist. The model effectively represents an all-to-all coupled spin-$1/2$ chain subjected to periodic kicks~\cite{olsacher2022digital}. It has been previously established that the system exhibits a robust $2$-DTC phase~\cite{14}, where regular behavior is indicated by the mean level-spacing ratio. We find that the system supports both $2$-DTC and dynamical freezing (DF) phases around alternating rotationally symmetric points. In the DF phase, the average magnetization remains close to its initial value for all times. Most interestingly, we observe $4$-DTC phases within the regular regime of the system for higher values of angular momentum. This phenomenon cannot be explained by the $\mathbb{Z}_2$ symmetry of the model. Instead, we interpret the emergence of the $4$-DTC phases using the semiclassical phase-space structure, specifically through the phase space bifurcations within a parametric regime. The effects of bifurcations are also present in the quantum dynamics of magnetization. 

The stability of the $4$-DTC phases is further analyzed using the linear entropy~\cite{9} for different angular momentum values. Moreover, we demonstrate the emergence of dynamical conservation for both the $2$-DTC and DF phases through the time behavior of the out-of-time-ordered correlator (OTOC)~\cite{swingle2018unscrambling,sreeram2025periodicity}.In contrast, the OTOC analysis does not reveal any dynamical conservation associated with the $4$-DTC phases.
The metrological applications of dynamical phases also have been demonstrated through the behavior of the QFI in the context of parameter estimation of the system. We have shown that the boundary regions of various dynamical phases helps in estimating parameters with higher precision compared to regions deep inside the phases. We also show that the boundary regimes of the 4-DTC phase provide highly sensitive quantum sensing platforms for the precise estimation of the rotational parameter.

The paper is organized as follows. In Sec.~\ref{sec:level2}, we describe quantum kicked top model and discuss its quantum and classical dynamics. The phase diagram separating the regular and chaotic regimes of the system is presented in Sec.~\ref{mlsr}, using the random-matrix-theory values of the mean level-spacing ratio. Various dynamical phases of the system, including the 4-DTC phase, are discussed in Sec.~\ref{DTC_DF}. To further investigate the correlation properties of the dynamical phases, we study the time dynamics of the linear entropy and the OTOC in Sec.~\ref{entropy_otoc}. The metrological advantages of the system, in the context of our results, are discussed in Sec.~\ref{metrology}. Finally, in Sec.~\ref{summary}, we summarize our results.

\section{\label{sec:level2}Model}
We consider an all-to-all coupled spin-$1/2$ model in a transverse field with periodic kicks that can also be defined as quantum kicked top (QKT). This is a paradigmatic model of quantum chaos that has been studied extensively in the context of quantum chaotic dynamics and more recently for digital quantum simulation. There are many variants of the QKT  with slight modifications \cite{2,5,7,8}, we consider the basic one introduced in Ref. \cite{1} for studying quantum chaos. 

\begin{equation}
    H(t) =\frac{p}{\tau}J_y + \frac{k}{2j} J_z^2 \sum_{n=-\infty}^{\infty}\delta(t-n\tau),
    \label{QKT_model}
\end{equation}
where $\{J_x, J_y, J_z\}$ are three components of an angular momentum having commutation relation $[J_i,J_j]=i\epsilon_{ijk}J_k$.
The Hamiltonian describes an angular momentum vector precessing around the y-axis with an angular frequency $p/\tau$ and going through state-dependent twists about the $z$-axis with periodic kicks in the interval $\tau$ and the strength $k$. The system has been experimentally realized using nuclear spins in a magnetic field in presence of laser pulses \cite{13}. The angular momentum $j$ of the system can be considered as the composition of $N=2j$ spin-$1/2$'s with $\textbf{J}=\sum_{i=1}^{N}\textbf{s}_i$. Thus, the Hamiltonian in Eq.~(\ref{QKT_model}) represents an infinite-range Ising model in transverse field having periodic kicks associated with the interaction term. Consequently, the model has a well established many-body description with microscopic degrees of freedom as spin-$1/2$. We explore this property to calculate linear entropy and to understand interesting dynamics of the system.

Due to the periodic nature of the drive, we can construct a Floquet operator that determines the evolution of the system over one time period which is used to study the dynamics of the system in stroboscopic times. The Floquet operator for our system is given by,
\begin{equation}
U = e^{-i\frac{k}{2j}J_z^2} e^{-ipJ_y} ,
\label{Floquet_op}
\end{equation}
where we consider $\tau=1$. Since $[J^2,H(t)]=0$, $\langle\psi(t)| J^2|\psi(t)\rangle=j(j+1)$ is a conserved quantity with $j$ being the total angular momentum. Therefore, the dynamics is confined to a single $J$-sector and the Hilbert dimension is given by $d=2j+1$ with the basis states $|j,m\rangle$, where $m$ is the projection of the angular momentum on $z$-axis. Due to the conservation of total angular momentum, the dynamics of $\{J_x, J_y, J_z\}$ is confined on the surface of a three-dimensional sphere. The time evolution of the angular momentum operators at stroboscopic time is given by
\begin{equation}
J_i^{'}(t+\tau)=U^{\dagger}J_i(t)U,
\end{equation}
where $i=x, y, z$. The classical limit of the quantum top can be obtained by considering $j\rightarrow\infty$. The classical map corresponding to the quantum kicked top (\ref{QKT_model}) is given as follows

\begin{eqnarray}
X^{\prime} &=& (X \cos p + Z \sin p) \cos\left(k\left(Z\cos p - X \sin p\right)\right)\nonumber\\ 
&& -Y \sin\left(k\left(Z \cos p- X \sin p\right)\right),\nonumber\\
Y^{\prime} &=& (X \cos p + Z \sin p)\sin \left(k\left(Z \cos p-X \sin p\right)\right) \nonumber\\ 
&& +Y \cos\left(k\left(Z \cos p-X\sin p\right)\right),\nonumber\\
Z^{\prime} &=& -X \sin p + Z \cos p,
\label{classical_map}
\end{eqnarray} 
where $X=J_x/j$, $Y=J_y/j$ and $Z=J_z/j$. The dynamical variables here follow the relation $X^2+Y^2+Z^2=1$ that indicates the dynamics is restricted on the unit sphere. This leads to the parametrization of the variables as $X=\sin\theta\cos\phi$, $Y=\sin\theta\sin\phi$ and $Z=\cos\theta$, where $\theta$ and $\phi$ are polar angle and azimuthal angle, respectively. As the system evolves following the Eq.~(\ref{classical_map}), the trajectories of the angular momentum vector can be represented in $(\theta,\phi)$ plane. In order to explore the quantum dynamics of the top, the most suitable choice of the initial state is SCS that is defined as
\begin{equation}
\ket{\theta,\phi} =e^{i\theta(J_x sin\phi -J_y cos\phi)}\ket{j,j},
\label{scs}
\end{equation}
This is also useful in investigating the quantum-classical correspondence of the kicked top in the context of quantum chaos. In this work, we consider both spin-polarized states and general SCSs to investigate the DTC and DF phases of the QKT, along with the associated correlation dynamics and metrological applications.

\section{Results}
\label{results}
Here we consider the dynamics of the system (\ref{QKT_model}) in various parametric regimes. Depending on the behavior of time-evolution of the average $z$-component angular momentum, we propose different dynamical regimes in the parameter space.

\subsection{Mean level spacing ratio}
\label{mlsr}
The energy-level statistics of a quantum Hamiltonian is one of the widely used measures of quantum chaos. For a  periodically driven system, the level or spectral statistics is analyzed through the quasienergies or eigenphases of the Floquet operator. The quasienergy spectrum of the Floquet operator is obtained as
\begin{equation}
U|\varPhi_i\rangle=e^{i\nu_i}|\varPhi_i\rangle,
\end{equation}
where $\nu_i$ is the $i$-th eigenphase of the Floquet operator $U$. The eigenphases are $2\pi$ periodic and we confine them between $-\pi$ to $\pi$. In this regard, the level spacing ratio \cite{3,19} is an important quantity to investigate since it does not depend on the local density of states and, therefore, does not require unfolding. The level spacing ratio is defined as
\begin{equation}
    r_n = \frac{\text{min}(s_n , s_{n-1})}{\text{max}(s_n , s_{n-1})},
    \label{level_ratio}
\end{equation}
where $s_n=\nu_{n+1}-\nu_n$ is the spacing between two consecutive quasienergy levels of the Floquet operator. Consequently, the mean level spacing is given by
\begin{equation}
    \langle r\rangle = \frac{1}{(d-2)}\sum_{n=1}^{d-1} r_n .
\end{equation}
Depending on the value of $\langle r\rangle$ one can state whether the system in hand is chaotic or integrable. For integrability, $\langle r\rangle\approx0.39$ that signifies the Poisson statistics of level spacings, implying the crossing of quasi energy levels. However, for the chaotic regime predictions of random matrix theory (RMT) is well suited which implies repulsion between any consecutive energy levels (no crossing). In the chaotic case $\langle r\rangle\approx0.53$ provided by the Wigner-Dyson statistics of level spacing.

\begin{figure}[htbp]
    \centering
    \includegraphics[scale=1, trim=50 0 0 0, clip=true]{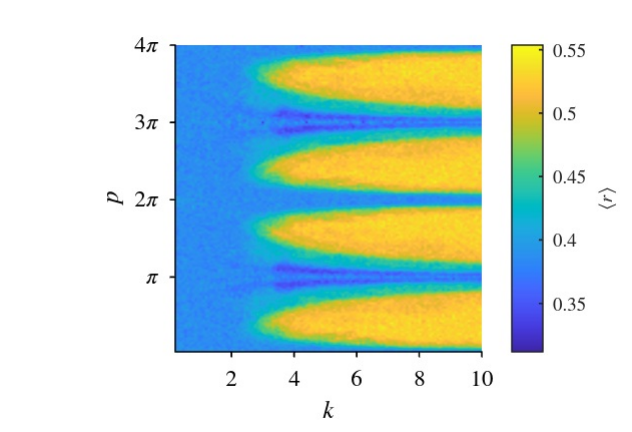}
    \caption{Variation of mean level spacing ratio as a function of $k$ and $p$. For $k\lesssim3$, the system shows regular behavior for any $p$. At integer multiples of $\pi$ in $p$, the system remains regular for any value of $k$. For other values of $p$, regular-to-chaotic crossover appears for $k\approx3$. Here, $j=1000$.}
    \label{fig_mlsr}
\end{figure}

In Fig. \ref{fig_mlsr}, we show a density plot of the mean level spacing ratio in the parameter space of $p$ and $k$. The system exhibits complete integrability for $k<3$, whereas, it becomes periodic between chaos and integrability as $p$ is changed by $\pi$ for $k>3$. This can be explained by the rotation symmetry $R_y=e^{-i\pi J_y}$ of the Floquet operator $U$ (\ref{Floquet_op}). In this case, $[U,R_y]=0$ as indicated by $J_x\rightarrow-J_x$, $J_y\rightarrow J_y$ and $J_z\rightarrow -J_z$ under the rotation operator $R_y$. As a result, the system becomes regular around $p=n\pi$, $n$ being integers for any kicking strength $k$.

\subsection{DTC and DF phases}
\label{DTC_DF}
A DTC is a non-equilibrium phase of matter that occurs as a consequence of a physical observable breaking the discrete time-translation symmetry of the many-body Hamiltonian. More precisely, there must exist an observable $\hat{O}$ and a class of initial states $\{|\psi\rangle\}$, so that the expectation value $\langle\hat{O}(t)\rangle=\text{lim}_{N\rightarrow\infty}\langle \psi(t)|\hat{O}|\psi(t)\rangle$ satisfies following three conditions: (I) Time-translation symmetry breaking: $\langle\hat{O}(t+\tau)\rangle\ne\langle\hat{O}(t)\rangle$ given $H(t)=H(t+\tau)$. (II) Rigidity: The observable oscillates with a time period $\tau_o=n\tau$ with $n(\ne1)$ being an integer without fine tuned system parameters. (III) Persistence: The oscillation sustains for an infinitely long time.

As mentioned before, our system can be assumed consisting of $N_{s}$ spin-1/2 with $N_{s}/2=j$. We can then define average angular momentum per spin as
\begin{align}
\expval{s_i} = \frac{\expval{J_i}}{N_{s}},
\end{align}
where $i=x,y,z$. In this work, we consider the average $z$-component of angular momentum or average magnetization, defined as $\langle m_z(n)\rangle=\frac{\langle J_z(n)\rangle}{N_s}$, for finding different dynamic phases of the system. Given regular and chaotic regimes of the system in parameter space (see Fig.~\ref{fig_mlsr}), we further investigate time-dependent behavior of average magnetization in different parametric regimes and observe various dynamical phases of the system.

\begin{figure}[htbp]
    \centering
    \includegraphics[width=0.51\textwidth, trim=0 0 0 17, clip=true]{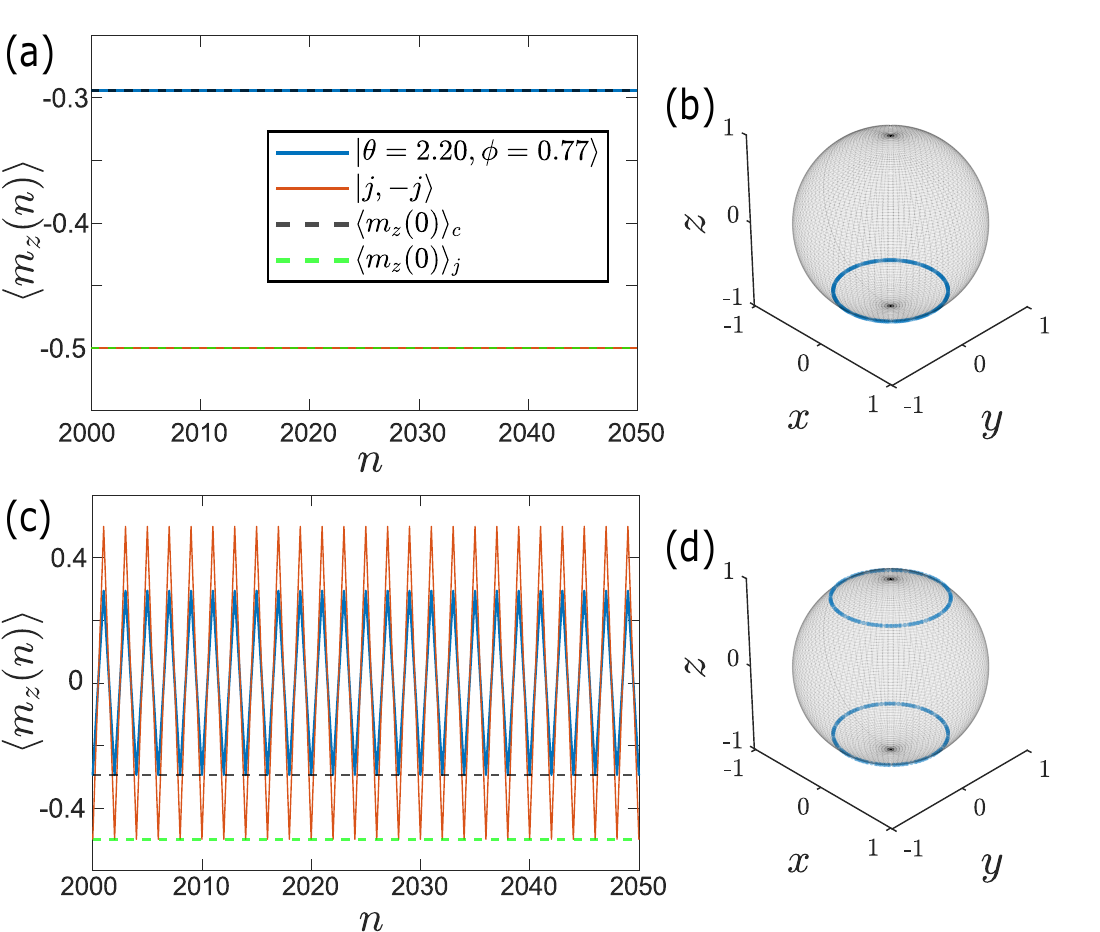}
    \caption{(a) Variation of the average magnetization, $\langle m_z \rangle$, with stroboscopic time for $p = 2\pi$ and $k = 6$ for two different initial states (spin polarized and SCS), illustrating dynamical freezing (DF). (b) The corresponding Poincaré surface plot obtained from classical dynamics for one initial condition and $300$ time steps with $k=6$. (c) Same as (a) but for $p = \pi$, exhibiting the $2$-DTC phase. (d) The corresponding Poincaré surface plot for the $2$-DTC case, showing transitions between two degenerate states. For both (a) and (c), the initial value of $\langle m_z(0) \rangle$ is shown throughout the time evolution as a reference.}
    \label{fig_dtc_dmf_am}
\end{figure}

By analyzing time-evolution of average magnetization, we find four dynamical phases: $2$-DTC, $4$-DTC, DF and a phase with decaying oscillation eventually zero magnetization in a long time limit. The last one is effectively showing chaotic behavior where the system thermalizes providing zero magnetization with some fluctuations in long-time limit. 
The $2$-DTC phase in the quantum kicked top has also been investigated previously in the context of a general $p$-spin all-to-all interacting spin-$1/2$ system, where it was proposed that the order ($q$) of the subharmonic response follows the relation $q \leq p$~\cite{14}. However, remarkably, we find $4$-DTC phases in the quantum kicked top model, which effectively represents an all-to-all interacting spin-$1/2$ system with $p=2$. In addition, we find DF phases where the average value of the magnetization remains close to the initial values forever. The existence of $4$-DTC and DF phases in the quantum kicked top model is not reported earlier to the best of our knowledge.

First, we consider an initial state that provides an exact expression of time-evolution of the average angular momentum. Thus, we choose an initial state, $|\psi(0)\rangle=\ket{j,-j}$ that provides $\langle m_z(0)\rangle=-0.5$ and $\langle m_z(n)\rangle=-0.5\cos np$, for $p=m\pi$ with $m$ being an integer. We can get two cases here, for even $m$, $\langle m_z(n)\rangle=-0.5$ whereas, for odd $m$, $\langle m_z(n)\rangle=-0.5\times(-1)^n$. The first case is shown in Fig.~\ref{fig_dtc_dmf_am}(a), where the average angular momentum remains at its initial value. We also consider a general spin coherent state with $\theta=2.2$ and $\phi=0.77$, and find the system remains close to the initial state. This phenomenon is called dynamical freezing. In the present case (see Fig.~\ref{fig_dtc_dmf_am}(a)), the time-evolved magnetization remains exactly at its initial value due to the fine-tuned parameters. However, it may happen that the system freezes to another state close to the initial one causing a small deviation from the initial magnetization with fluctuations/beating-like structure around the mean value in other parameter regimes. To observe the behavior of classical trajectories for the case of DF, we investigate Poincar\'e surface for $p=2\pi$ for one initial condition (see Fig~\ref{fig_dtc_dmf_am}(b)). We can clearly see that the system is confined to a closed trajectory where $\theta$ remains fixed, leading no change in magnetization during the evolution. We have also checked the result by a small but finite deviation from $p=2\pi$, and found the same result, thus justifying the robustness of the DF phase. The second case with $m=1$, i.e., $p=\pi$ is shown in Fig.~\ref{fig_dtc_dmf_am}(c). The magnetization shows periodic oscillation with period $2\tau$, thus breaking the time translation symmetry of the Hamiltonian. We consider here two initial states: $|j,-j\rangle$ and spin coherent state $|\theta,\phi\rangle$ with $\theta=2.2$ and $\phi=0.77$. For both cases, the average magnetization shows exact periodic oscillations with the same period $2\tau$, but different amplitude. This indicates the existence of DTC around $p=\pi$. The same behavior is observed on Poincar\'e surface (see Fig.~\ref{fig_dtc_dmf_am}(d)), where the dynamics for one initial condition is confined between two closed trajectories at some $\theta$ and $\pi-\theta$ having two average magnetization $M_z$ and $-M_z$.\\
\begin{figure}[htbp]
    \centering
     \includegraphics[width=0.495\textwidth]{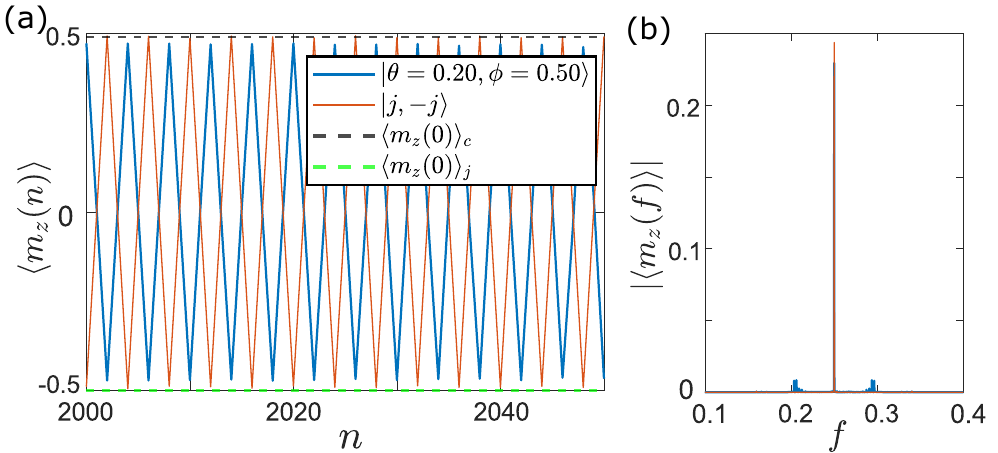}
    \caption{(a) Variation of $\langle m_z(t) \rangle$ with stroboscopic instant $n$, showing $4$ periodic subharmonic response for $k = 1.5$ with $p = \pi/2$ and $J = 100$. Two initial states, $\ket{j, -j}$ and $\ket{\theta = 0.2, \phi = 0.5}$, are considered. (b) Corresponding Fourier transform of $\langle m_z(t) \rangle$.}
    \label{fig_4dtc_am}
\end{figure}

\begin{figure*}[htbp]
    \centering
    \includegraphics[width=1.05\textwidth]{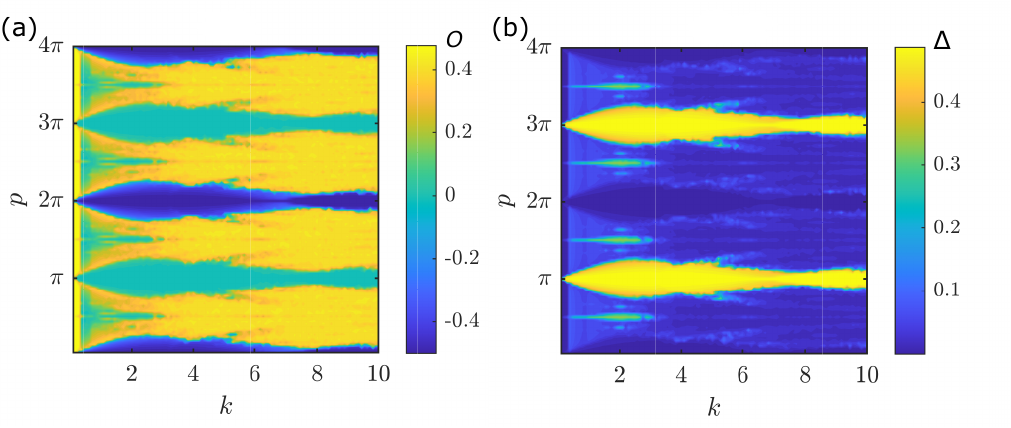}
    \caption{Density plots of (a) the order parameter ($O$) and (b) the standard deviation ($\Delta$) for $J = 100$ after $4000$ kicks, considering $\ket{j, -j}$ as the initial state. The order parameter distinguishes the DTC ($O \approx 0$), DF ($O \approx -0.5$), and chaotic ($O \approx 0.5$) phases, whereas $\Delta$ captures time-crystalline behavior ($\Delta \neq 0$) and vanishes ($\Delta = 0$) in both the chaotic and DF phases. The values of $O$ and $\Delta$ for different phases correspond exclusively to the initial state $\ket{j, -j}$.}
    \label{fig_op_os}
\end{figure*}

We are now interested to examine the robustness of DTC and DF behavior in the context of average magnetization.
For a proper distinction of different dynamical phases, we construct a non-conventional order parameter, defined as

\begin{align}
O =[ {\rm Sgn}(\cos \theta) (\expval{m_z(0)}-\overline{\expval{m_z}}) -|\expval{m_z^{max}}|] ;\nonumber \\ \hspace{2mm} \overline{\expval{m_z}}=  \frac{1}{N} \sum_{n=1}^{N} \langle m_z(n)\rangle,
\label{op1}
\end{align}
where $N$ is the number of stroboscopic time steps. The overhead bar denotes time averaging, whereas $\langle...\rangle$ indicates quantum average. The quantity, $|\expval{m_z^{max}}|$ is absolute maximum value of $\langle m_z(n)\rangle$ between $N/2$ and $N$ stroboscopic instants. For both DTC and thermalized phases, the time averaged magnetization is zero. Therefore, this is not a good order parameter to differentiate the dynamical phases of the system. On the other hand, the order parameter proposed in Eq.~(\ref{op1}) is capable of successfully defining various dynamical phases with different numerical values that depend on the initial quantum states. We describe here the values of $O$ for different phases in context of the initial state $|j,-j\rangle$, however, this order parameter is valid for any general SCS with different numerical values for all the phases (see Appendix~\ref{Appendix:B}). In this case, the initial value of the average magnetization, $\langle m_z(0)\rangle=-0.5$. The quantity $|\langle m_z^{\text{max}}\rangle|=0.5$ for both DTC and DF phases, whereas it is $0$ for the thermal (chaotic) phase. On the other hand, $\overline{\expval{m_z}}=0$ for both DTC and thermal phases, and it is $-0.5$ for the DF phase. Using these results, we find that the order parameter $O=-0.5, 0$ and $0.5$ for DF, DTC and thermal phases, respectively for the initial state $|j,-j\rangle$. We have shown a density plot for the order parameter $O$ by varying the parameters $k$ ans $p$ in Fig.~\ref{fig_op_os}(a). We clearly observe the $2$-DTC phase around $p = m\pi$, where $m$ is an odd integer, for any value of $k$, whereas the DF phase appears for $p = m\pi$, where $m$ is an even integer, for all values of $k$. These dynamical phases persist over a finite region of the parameter space, thereby demonstrating their robustness. Again, if we change the initial state to an arbitrary SCS, the phase diagram in Fig.~\ref{fig_op_os}(a) will qualitatively remain the same; only the value of the order parameter will differ.

The observed symmetry of the order parameter around $p$ originates from the specific choice of a $Z_2$-symmetry-broken state; for other states with finite $z$-magnetization, this symmetry may not persist (see Figs.~\ref{fig_op_os} and~\ref{fig_correspondence}). A careful inspection of the order-parameter density plot shows finite positive values of the order parameter around $p=\pi/2$, indicating the possible emergence of a higher-order response. To investigate this possibility, we focus on the fluctuations in the time variation of magnetization across the parameter space. The expectation is that this measure can serve as an indicator of the existence of a higher-order DTC phase. We define the standard deviation of $\langle m_z(n)\rangle$ as

\begin{align}
\Delta = \sqrt{ \frac{1}{N-1}\sum_{n} (\expval{m_z(n)} -\overline{\expval{m_z}})^2},
\end{align}
where $N$ denotes total stroboscopic time steps. We find that $\Delta$ becomes zero for both the DF and chaotic phases, whereas it takes finite, nonzero values for the DTC phases. For the initial state $|j, -j\rangle$, $\Delta$ takes the value $0.5$ in the $2$-DTC phase, whereas $\Delta < 0.5$ indicates the presence of a higher-order DTC phase. We show a density plot of $\Delta$ as a function of $k$ and $p$ in Fig.~\ref{fig_op_os}(b) for the initial state as $|j,-j\rangle$. In addition to the $2$-DTC phases shown in Fig.~\ref{fig_op_os}(a), we find higher-order DTC phases for $p = m\pi/2$, where $m$ is an odd integer. As we will see below, these are indeed $4$-DTC phases that appear for higher values of $J$.

\begin{figure*}[htbp]
    \centering
     \includegraphics[width=1.12\textwidth]{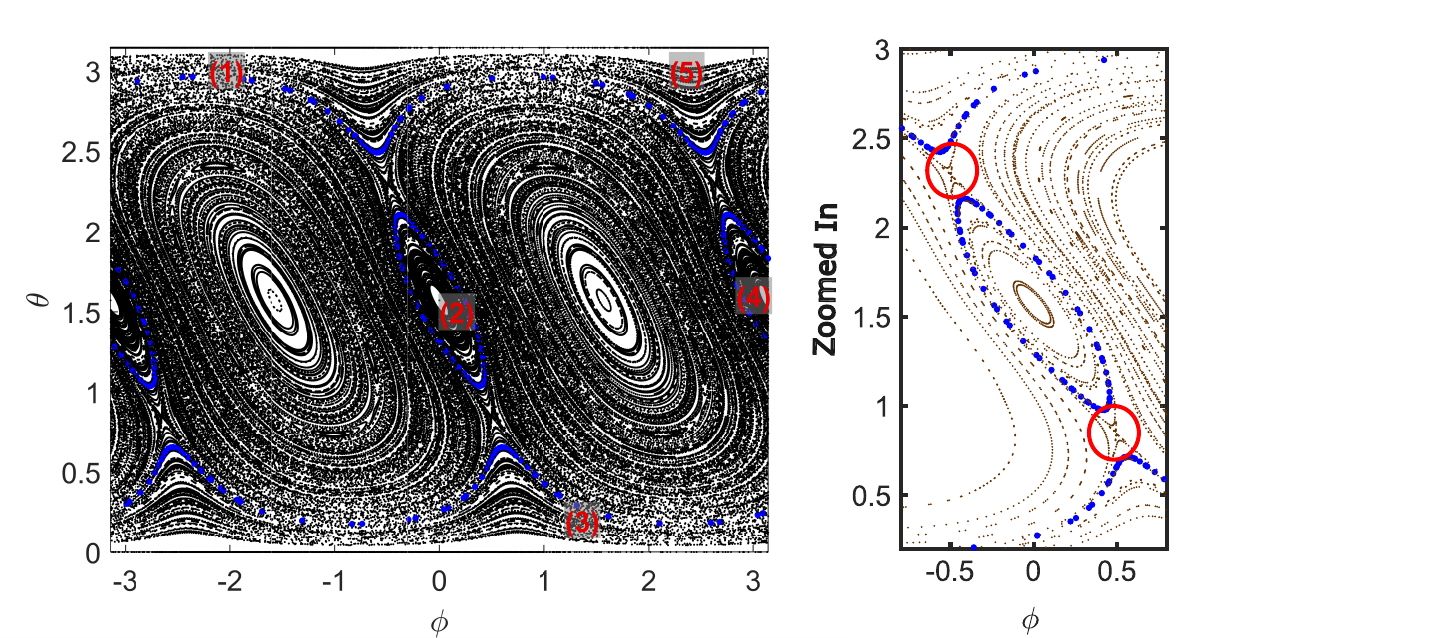}
    \caption{ Two-dimensional phase-space plot in $\theta-\phi$ plane for $k=1.5$ and $p=\pi/2$. A special island is formed around $\theta=\pi/2$ and $\phi=0$ as indicated by point $(2)$ in the plot.  For the $4$-DTC phase (see Fig.~\ref{fig_4dtc_am}, the initial state around $\theta \approx \pi$ visits the four islands marked with the numbers in red. In the right side, the zoomed in picture around point $(2).$}
    \label{fig_phasespace}
\end{figure*}

We now focus on the time evolution of the average magnetization around the $4$-DTC phases, as shown in Fig.~\ref{fig_op_os}(b). The existence of the $4$-DTC phase is not intuitive for such a simple system possessing $\mathbb{Z}_2$ symmetry. We also observe that the $4$-DTC phase emerges for higher values of $J$, specifically for $J > 20$ in our case. A similar behavior has been reported in Ref.~\cite{27}. An exact periodic oscillation of the average magnetization with a time period of $4\tau$ is observed in Fig.~\ref{fig_4dtc_am}(a) for $J = 100$, $k = 1.5$, and $p = \pi/2$, and the corresponding Fourier transform is shown in Fig.~\ref{fig_4dtc_am}(b).

It is important to note that while the $2$-DTC phase has a proper quantum description, namely 
$[U,R_y]=0$ and hence the eigenstates of $R_y$ are simultaneously the eigenstates of $U$, leading to $\pi$-paired eigenstates of $U$, the $4$-DTC phase does not possess an analogous $\pi/2$ eigenphase pairing structure.
To investigate the origin of the $4$-DTC phase, we consider the classical dynamics of the system. There is also evidence for the semiclassical origin of the $4$-DTC phase, as it appears only for $J>20$ in our system. The phase plot in the $\theta$–$\phi$ plane is shown in Fig.~\ref{fig_phasespace} for the same values of $k$ and $p$ considered in Fig.~\ref{fig_4dtc_am}. We observe an interesting phase-space dynamics for $k = 1.5$ and $p = \pi/2$, which explains the existence of the four-periodic DTC phase in the system (see Fig.~\ref{fig_phasespace}(a)). Following a close inspection of the phase-space plot, we observe the formation of an island that acts as a mediator between the two flows along $\theta=\pi$ and $\theta=0$ (see the red circles in the zoomed-in plot, which highlight the formation of “bridges”). In the scenario where the initial state is chosen near $\theta=\pi$, marked by $(1)$, the system reaches the point $5$ with the same initial value of $\theta$ after traversing through the points $(2), (3) {\rm and} (4)$. Since the $z$-magnetization depends solely on $\theta$, the system exhibits $4$-DTC behavior, as illustrated in in Fig.~\ref{fig_4dtc_am}(a). It can be noted that the $4$-DTC behavior persists for initial states chosen anywhere within the marked regime, allowing for a general coherent state characterized by finite $\theta$ and $\phi$. The observed $4$-DTC behavior is not restricted to a fine-tuned point, but persists over a finite parameter regime, as indicated in Fig.~\ref{fig_op_os}(b). In Appendix~\ref{Appendix:A}, we present the phase-space dynamics and the quantum dynamics of the average $z$-magnetization as functions of the parameter $k$, with $p$ fixed at $\pi/2$ (see Fig.~\ref{fig_bifurcation}). The results exhibit four-fold bifurcations in $\theta$ and in the average $z$-magnetization for the classical and quantum dynamics, respectively, over a range of $k$ where the $4$-DTC phase emerges following a chaotic regime. We also observe that this range of $k$ depends on the initial state, consistent with the behavior observed for the $4$-DTC phase. In Appendix~\ref{Appendix:A}, we also present an analytical calculation based on the classical map to explain why the initial states with $\theta=0$ or $\pi$ exhibit $4$-DTC behavior.

\subsection{Linear entropy and OTOC}
\label{entropy_otoc}
In this section, we examine the dynamics of the linear entropy and the OTOC for the different dynamical phases observed in this work and explain their signature. The linear entropy (LE) is obtained as a linearized form of the more general von Neumann entropy.
As mentioned earlier, the total angular momentum $J$ can be regarded as $N_s = 2J$ spin-$\tfrac{1}{2}$ particles. The entropy can thus be defined as the entanglement between a single spin and the remaining $N_s - 1$ spins. For our system, the linear entropy can be represented as~\cite{9}
\begin{align}
S = \frac{1}{2} - \frac{1}{2j^2}(\expval{J_x}^2 +\expval{J_y}^2+\expval{J_z}^2) .
\end{align}
This can be observed that the minimum and the maximum value of the LE are $0$ and $0.5$ for pure state and maximally mixed state, respectively. 

\begin{figure}[h!]
    \centering
     \includegraphics[width=0.5\textwidth]{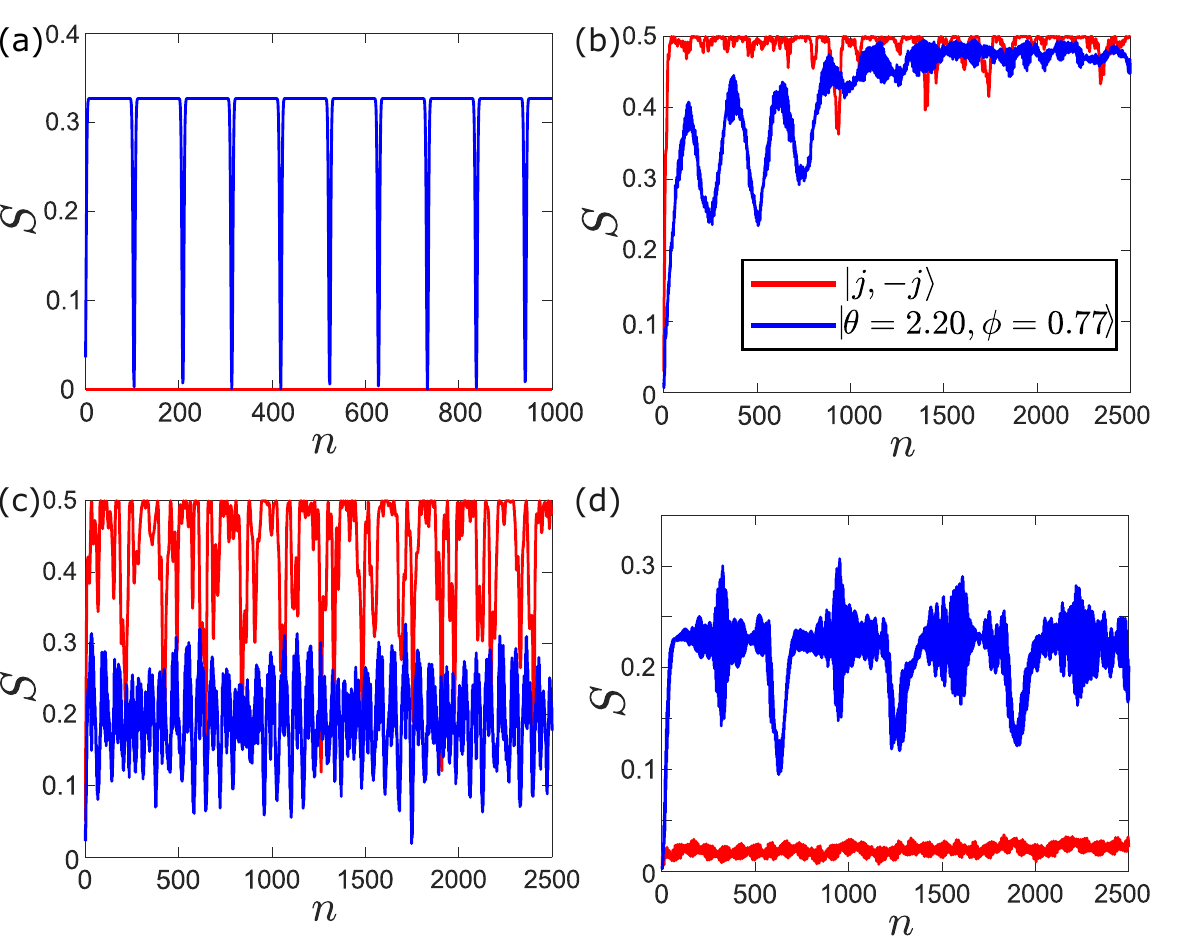}
    \caption{ Variation of linear entropy (LE) with stroboscopic time: (a) $p = \pi$, $k = 6$, $j = 100$; (b) $p = \pi/2$, $k = 3.5$, $j = 100$; (c) $p = \pi/2$, $k = 1.5$, $j = 10$; (d) $p = \pi/2$, $k = 1.5$, $j = 100$. Upper panel: (a) and (b) show the behavior of entanglement entropy in the DTC and chaotic regions, respectively. Lower panel: (c) and (d) demonstrate that the LE decreases with the emergence of higher-order DTC as the system size increases from $j = 10$ to $j = 100$.}
    \label{fig_le}
\end{figure}

We investigate the dynamics of the linear entropy when the system is confined to various dynamical regimes. For $p = \pi$ and $p = 2\pi$, the system exhibits the $2$-DTC and DF phases, respectively, for any value of $k$. We first evaluate the time evolution of the linear entropy considering the initial state $|j, -j\rangle$, which yields $\langle J_x(t) \rangle = \langle J_y(t) \rangle = 0$ and $\langle J_z(t) \rangle = -j$ for $p = \pi$. This results in zero linear entropy at all times (see Fig.~\ref{fig_le}(a)), which is also the case for $p = 2\pi$. However, for a general spin coherent state (SCS), we observe a finite entropy since all the components of angular momentum contribute in the dynamics. Interestingly, for any arbitrary initial state yielding finite magnetization, the entropy exhibits periodic maxima and minima over time when $p = \pi$ and $p = 2\pi$ (see Fig.~\ref{fig_le}(a)). For $p = \pi/2$ and $k = 3.5$, which corresponds to a chaotic phase, the linear entropy of the kicked top saturates around $0.5$ for both the $\ket{j, -j}$ state and a general SCS, as shown in Fig.~\ref{fig_le}(b). We also evaluate the LE for different values of $j$ in the DF, $2$-DTC, and chaotic phases (not shown). The overall behavior remains similar; however, the periodicity may vary in the DF and $2$-DTC phases, and the saturation time differs in the chaotic case. Most remarkably,
we find that, as the value of $J$ increases within the $4$-DTC phase, the linear entropy decreases, thereby stabilizing the corresponding phase (see Figs.~\ref{fig_le}(c,d)). Therefore, there exists a one-to-one correspondence between the reduction of linear entropy and the emergence of higher-order DTC phases for the large value of $J$.

\begin{figure*}[htbp]
    \centering
     \includegraphics[width=1.1\textwidth]{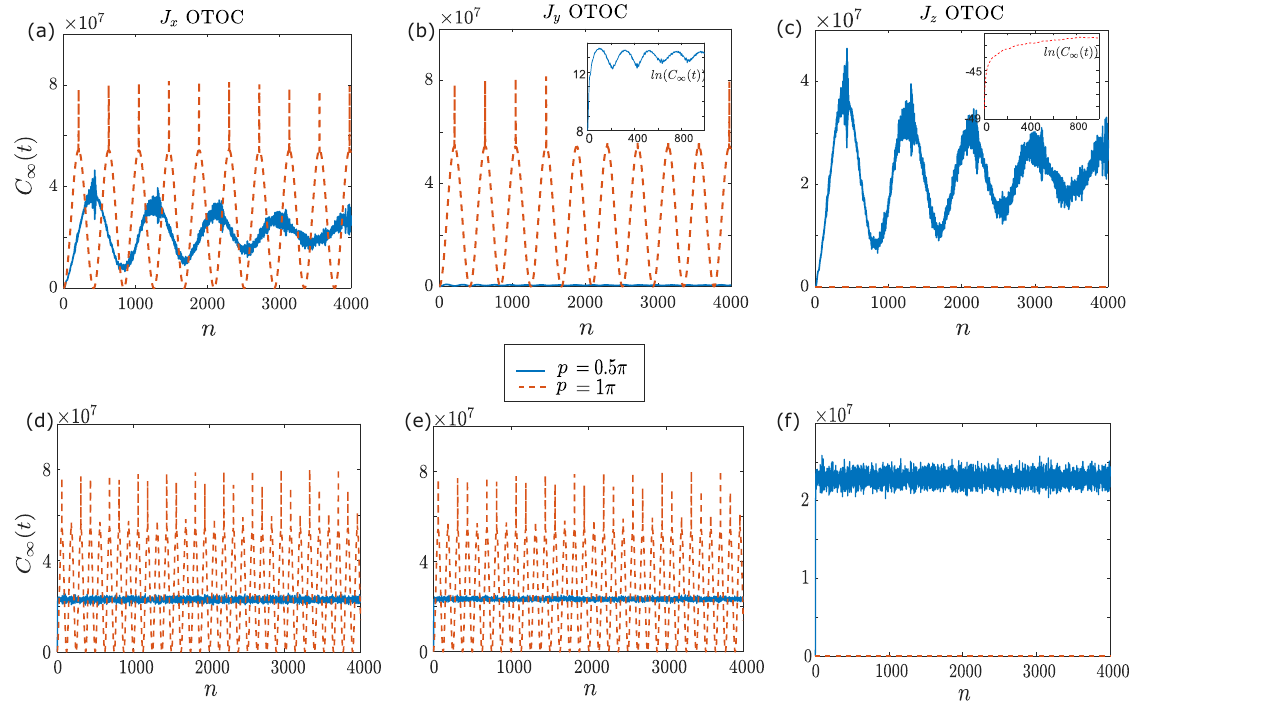}
    \caption{Evolution of the OTOC corresponding to the $J_x$, $J_y$, and $J_z$ operators with stroboscopic time, as indicated in the plots. The upper panel corresponds to $k = 1.5$, whereas the lower panel shows results for $k = 5$. The infinite-temperature OTOC remains zero at all times for $J_z$ in the case of $p = \pi$ (and also for $p = 2\pi$, not shown here), demonstrating the emergence of dynamical conservation in the DTC and DF phases. However, for $p = \pi$, the OTOC corresponding to $J_x$ and $J_y$ exhibits a periodic recurrence of dynamical conservation. For $p = \pi/2$ and $k = 5$, the OTOC in all cases saturates at higher values, indicating the onset of quantum chaos. Here, $j = 100$.}
    \label{fig_otoc}
\end{figure*}

The out-of-time-ordered correlator (OTOC) serves as a useful measure of quantum chaos and integrability. In general, the OTOC initially exhibits exponential growth with time and eventually saturates to a large value in the chaotic regime, whereas it displays oscillatory behavior in the regular regime before saturating to a lower value than in the chaotic case~\cite{10,11}.
The OTOC is typically defined by
\begin{align}
\begin{split}
C_T(t) = \left\langle [A(t), B(0)]^\dagger \, [A(t), B(0)] \right\rangle,
\end{split}
\end{align}
where $\expval{...}$ is thermal average and A, B are observables of the system. We study infinite-temperature OTOC and consider A(t) as $J_i(n\tau)$ and B(0) as $J_i(0)$. In our case, this infinite-temperature OTOC can be written as
\begin{align}
C_\infty(n)=-\frac{1}{d}\ Tr([J_i(n), J_i(0)]^2).
\end{align}
Our definition of the OTOC is independent of the initial state. 

We find that for $p = \pi$ and $p = 2\pi$, the commutator $[J_z(n\tau), J_z(0)] = 0$, since $J_z(n) = (-1)^nJ_z(0)$ or $J_z(0)$ depending on whether $p$ is an odd or even multiple of $\pi$ (corresponding to the $2$-DTC and DF phases, respectively). This result implies the conservation of $J_z$ at all times (see Figs.~\ref{fig_otoc}(c,f)). Interestingly, even though the Hamiltonian $H(t)$ does not commute with $J_z$, the dynamics within this parameter regime enforce an effective conservation of $J_z$. Such behavior represents dynamical conservation, meaning that the conservation law emerges purely due to the system’s dynamics.

Furthermore, the OTOCs of $J_x$ and $J_y$ exhibit periodically vanishing commutators, indicating the periodic emergence of dynamical conservation for these components as shown in Figs.~\ref{fig_otoc}(a,b,d,e). However, for $p = \pi/2$, the OTOCs of $J_x$ and $J_z$ display similar temporal behavior but remain nonzero, unlike in the DF or $2$-DTC phases. In this case, the OTOC of $J_y$ attains a comparatively small value (approximately $10^5$) relative to the maximum OTOC value (approximately $10^7$). Although $4$-DTC is manifested around $p = \pi/2$ and $k = 1.5$, no dynamical conservation is observed. As expected, in the chaotic regime the OTOC grows rapidly and eventually saturates to a finite value with time (see plot for $p=\pi/2$ in Fig.~\ref{fig_otoc}(f) ).

\section{Metrological Benefits}
\label{metrology}

\begin{figure*}[htbp]
    \centering
    \includegraphics[width=0.95\textwidth, trim=0 0 0 0, clip=True]{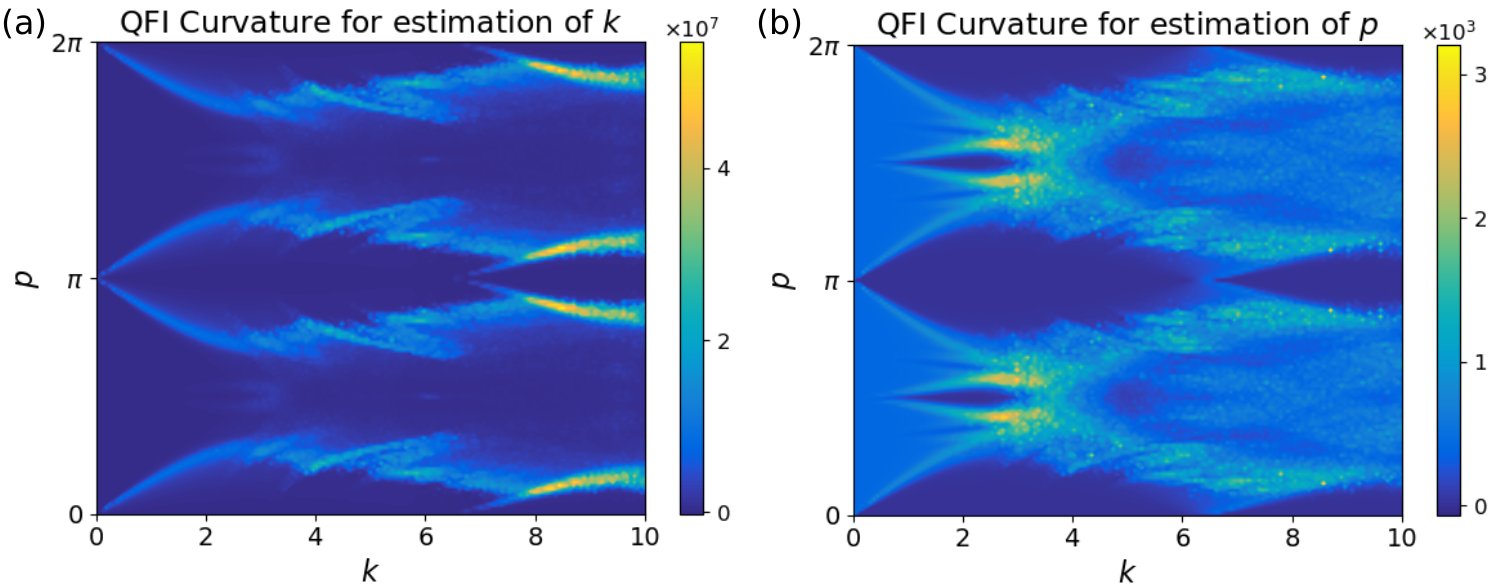}
    \caption{ Density plot of QFI Curvature for estimation of the parameter (a) $k$ and (b) $p$ . Here, $j = 100$ and $t=500$.}
    \label{fig_qfic}
\end{figure*}
Having established and characterized the various dynamical phases of the system, we now focus on investigating their effectiveness for metrological applications. The quantum phase transitions play a crucial role in enabling high-precision estimation of various system parameters. This is known as critical quantum metrology, where the regions near the phase transition points exhibit immense sensitivity to the system parameters.

\subsection{Quantum parameter estimation using QFI}
The main aim of quantum parameter estimation theory is to estimate an unknown parameter (say $\lambda$) encoded in a quantum state $\rho(\lambda)$.
Typical measurements are described by Positive Operator-Valued Measures (POVMs), where the probability of obtaining the $s$-th outcome is given by $p_{s}={\rm Tr}(\rho(\lambda)\Pi_{s})$, where $\Pi_{s}$ is the POVM operator corresponding to the $s$-th outcome.
For repeated quantum measurements performed on the probe state, the measurement outcomes yield different estimates of the parameter $\lambda$. The variance of these estimates is lower bounded by the quantum Cramér-Rao bound,
\begin{align}
    (\Delta\lambda)^{2}\ge \frac{1}{mF_{Q}(\lambda)},
\end{align}
where m is the number of repeated measurements and $F_{Q}(\lambda)$ is the Quantum Fisher Information (QFI). The QFI sets the ultimate precision bound achievable through optimization over all possible POVMs. For pure quantum states, where $\rho(\lambda)=|\psi(\lambda)\rangle\langle\psi(\lambda)|$, the QFI can be evaluated as
\begin{align}
F_Q(\lambda)=4 [\langle\partial_{\lambda} \psi(\lambda)|\partial_{\lambda} \psi(\lambda)\rangle-|\langle\psi(\lambda)|\partial_{\lambda} \psi(\lambda)\rangle|^2]. 
\end{align}

Since the system evolves under the Floquet operator, the quantum state encodes the parameter at every stroboscopic time step. Consequently, the QFI evolves with time and may exhibit significant enhancement in certain dynamical regimes. This growth in QFI can be characterized through the QFI curvature, which is simply defined as the second derivative of the QFI with respect to time.
\begin{equation}
    \kappa_{\lambda}=\frac{d^{2}F_{Q}(\lambda)}{dt^{2}}, \ \ \ \lambda \in \{p,k\}.
\end{equation}
We consider the QFI curvature, $\kappa_{\lambda}$, since the quantum Fisher information $F_Q(\lambda)$ exhibits an approximate quadratic growth with time, thereby providing a meaningful measure of the sensitivity of the system with respect to variations in the parameter $\lambda$ (see Appendix \ref{Appendix:B}). In Ref.~\cite{Ullah2026}, the metrological advantages were demonstrated for a few-body system. Here, we consider a large value of $j$, which effectively corresponds to a system containing a large number of spin-$1/2$.

\subsection{Estimation of $p$ and $k$}

To estimate a parameter, in general, one must choose an initial probe state such that the corresponding generator of the parameter translation exhibits a large variance. For the estimation of the kicking strength $k$, the relevant generator is $J_z^2$.Since the probe state $|j,-j\rangle$ is an eigenstate of $J_z$, the QFI nearly vanishes at the symmetry points $p=\pi$ and $2\pi$, while it attains a finite nonzero value at $p=\pi/2$. However, away from these symmetry points, the QFI generally exhibits finite quadratic time scaling, leading to a finite QFI curvature. Interestingly, although the QFI curvature cannot distinguish the different dynamical phases themselves, pronounced peaks appear near the boundaries between the phases (see Fig.~\ref{fig_qfic}(a)). These peaks indicate enhanced sensitivity around the dynamical phase-transition regions and therefore provide useful signatures for the estimation of the parameter $k$.

Interestingly, when estimating the parameter $p$, the corresponding generator is $J_y$, and the state $\ket{j,-j}$ is not an eigenstate of this generator. Consequently, the QFI or equivalently the QFI curvature, mostly distinguishes the different dynamical phases. In particular, the chaotic region naturally exhibits higher QFI values compared to the regions around the symmetry points, highlighting the potential usefulness of chaotic sensors \cite{Fiderer2018}. However, the QFI curvature is unable to clearly distinguish between the $2$-DTC and DF phases. As expected, the QFI becomes large near the boundaries between different dynamical phases, which is also evident in Fig.~\ref{fig_qfic}(b). More interestingly, the region around $p=\pi/2$ displays a QFI curvature larger than that of any other region. This occurs because the period-$4$ orbit represents the most sensitive part of the phase space and can be destabilized when either $k$ or $p$ is increased beyond an allowed regime. Therefore, the enhanced QFI curvature around $p=\pi/2$ originates from the presence of two possible phase transitions: one driven by increasing $k$, which pushes the dynamics into a chaotic regime, and the other driven by increasing $p$, which transforms the system from a $4$-DTC structure to a $2$-DTC or DF structure, with an intermediate region exhibiting unstable oscillations. To establish the genericity of the probe state dependence of our QFI results, we consider a general initial state $|\theta,\phi\rangle$ close to $|j,j\rangle$ and show that the QFI curvature exhibits qualitatively similar behavior, with only minor modifications (see Fig.~\ref{fig_correspondence} in Appendix~\ref{Appendix:B}). In the right set of plots in Fig.~\ref{fig_correspondence}, we consider an initial state that does not exhibit the $4$-DTC phase, and consequently, no enhanced sensitivity is observed around $\theta=\pi/2$. This confirms that the enhanced sensitivity originates from the presence of the $4$-DTC phase. In the same plots, we also show the correspondence between the order parameter and oscillation strength with the QFI curvature, revealing very good agreement among their respective behaviors.

 Thus, for a given dynamical system and probe state, the estimation of an encoded parameter is determined by the dynamics of the corresponding generator associated with the translation of that parameter. Since higher-order time crystals are highly sensitive to both the interaction and the choice of probe state, their experimental detection becomes extremely challenging \cite{zhang_expt_HO}. In this work, however, we demonstrate how higher-order time crystals can be identified for suitable probe states and appropriately defined order parameters. Furthermore, we characterize the sensitivity of these higher-order responses through the quantum Fisher information (QFI) curvature, which also validates the proposed order parameter.

\section{summary and conclusion}
\label{summary}
We have studied various dynamical phases of the quantum kicked top, by investigating the average magnetization, linear entropy, and OTOC, and have further demonstrated the metrological applications of these phases for the efficient estimation of physical parameters. The system exhibits both regular and chaotic regimes within its parameter space. It also possesses rotational symmetry for $p = n\pi$, where $n$ is an integer, and displays regular behavior for any value of the kick strength in these regions (see Fig.~\ref{fig_mlsr}). We have further classified the regular regimes of the system into various dynamical phases using our proposed order parameter.

The existence of a $2$-DTC phase around $p = \pi$ has been reported previously. In this work, we identify robust $2$-DTC and dynamical freezing (DF) phases around the odd and even multiples of $p$, respectively. The emergence of the $2$-DTC phase can be understood from the $\mathbb{Z}_2$ symmetry of the system, where it oscillates between two degenerate configurations. Interestingly, in this simple  model, we also observe higher-order discrete time crystal (DTC) phases, specifically, a $4$-DTC phase around $p = n\pi/2$, with $n$ being an odd integer. This phenomenon cannot be explained with the associated $\mathbb{Z}_2$ symmetry of the system.This result is in contrast to previous studies, where it was argued that the order of the DTC response in a $p$-body all-to-all interacting spin-$1/2$ system follows the relation $q \leq p$~\cite{14}. In our case, however, even with $p=2$, we observe a robust $4$-DTC phase.
The $4$-DTC phase emerges only for higher values of the angular momentum, as supported by the linear entropy results shown in Figs.~\ref{fig_le}(c,d). We find that the origin of the $4$-DTC phase is classical, in contrast to the $2$-DTC phase, and can be understood from the special islands that emerge in the phase space at $p=\pi/2$. We show that the $4$-DTC phase appears only for specific choices of initial states, whereas the $2$-DTC phase can be observed for arbitrary initial states. In experiments, the state preparation is often not perfectly precise. Nevertheless, we show that a $4$-DTC response can emerge in our system even without a specially prepared symmetry-broken initial state. Therefore, the possible experimental detection of $4$-DTC phase in our system is feasible.    We do not find any dynamical conservation law associated with the $4$-DTC phase. However, dynamical conservation is evident in both the $2$-DTC and DF phases, where we observe $[J_z(t), J_z(0)] = 0$ for arbitrary times. In the context of metrology, we find that the physical parameters of the system can be estimated with very high accuracy near the boundaries of the dynamical regimes by exploiting the QFI curvature. In particular, the boundary of the $4$-DTC phase can be used as a highly sensitive quantum sensor for estimating the rotational parameter or in experimental language the magnetic field.

In conclusion, we have identified higher-order DTC phases in one of the simplest quantum chaotic models, which is indeed an all-to-all coupled spin-$1/2$ chain driven by periodic kicks. We have also demonstrated the possible metrological applications of the dynamical phases. In particular, the boundary of the $4$-DTC phase can serve as an efficient quantum sensor for estimating the rotational parameter. A detailed study of the quantum sensing aspects will be presented elsewhere. In the near future, we aim to explore the possibility of realizing the proposed dynamical phases through qubit-based simulations.  Furthermore, we plan to investigate the effects of disorder and the breaking of permutation symmetry to examine the robustness of these dynamical phases.

\section*{Acknowledgement}
The authors acknowledge useful discussions with J. Bharathi Kannan, Andrea Pizzi and Arul Lakshminarayan. V. K. and A. R.  acknowledges support from IIT Hyderabad, India through the Seed Grant SG/IITH/F337/2023-24/SG-175. S. D. and A. R acknowledges support from  ANRF (DST), Govt. of India under the grant no. ANRF/IRG/2024/001653/PS.

\bibliography{Files/paper_ref}

\onecolumngrid
\appendix

\section{Emergence of $4$-period orbits at $p=\pi/2$}
\label{Appendix:A}
\begin{figure}[htbp]
    \centering
    \includegraphics[width=1.0\textwidth]{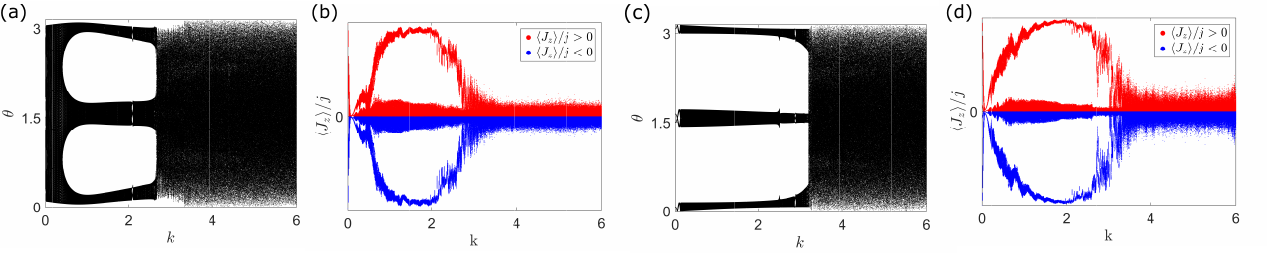}
\caption{(a) Variation of $\theta$ as a function of $k$ at $p=\pi/2$ with initial values $(\theta,\phi)=(0.2,0.5)$ in classical dynamics. (b) Average of $J_z(t)$ as a function of $k$ with an initial coherent state $|\theta=0.2,\phi=0.5\rangle$ and $p=\pi/2$ in quantum dynamics. (c) Same as (a) with $(\theta,\phi)=(1.5,0)$. (d) Same as (b) with $|\theta=1.5,\phi=0\rangle$. All the plots indicate signatures of the four-branch bifurcation at $p=\pi/2$ for different initial conditions. For quantum dynamics, $j=100$.
}
\label{fig_bifurcation}
\end{figure}

In this section, we explore the characteristics of the  semi-classical map in Eq.~\ref{classical_map} at the special points $p=\pi$ and $\pi/2$, where the $2$-DTC and $4$-DTC phases emerge, respectively.
The map can be written in the matrix form as,

\begin{align*}
M_{\pi/2} =
\begin{bmatrix}
0  & \sin kX & \cos kX \\
0 & \cos kX & -\sin kX \\
-1 & 0 & 0
\end{bmatrix};
\vspace{4cm}
M_{\pi}=
\begin{bmatrix}
-\cos kZ & \sin kZ & 0\\
\sin kZ & \cos kZ & 0\\
0 & 0 & -1
\end{bmatrix}.
\end{align*}
Interestingly, $M_{\pi}^2 = I$ for any value of $(X,Y,Z)$. This indicates the emergence of period-$2$ fixed points throughout the phase space. Thus, $2$-DTC appears near $p=\pi$ for an arbitrary initial state (see Fig.~\ref{fig_op_os}). On the other hand, at $p=\pi/2$, where the $4$-DTC phase emerges, $M_{\pi/2}^{4} \neq I$ for general values of $(X,Y,Z)$. However, there exists a region in phase space where $M_{\pi/2}^{4}=I$, which essentially indicates the emergence of period-$4$ orbits. We can write $M_{\pi/2}^4$ as
\[
\begingroup
\setlength{\arraycolsep}{3pt}
\renewcommand{\arraystretch}{1.2}
\resizebox{\linewidth}{!}{$
\begin{bmatrix}
\cos^2\alpha +0.5\sin(2\alpha)\sin(\alpha)
&
-0.5\cos(\alpha)\sin(2\alpha)
+0.25\sin(2\alpha)(1+\cos(2\alpha))
-0.5(1-\cos(2\alpha))\sin(\alpha)
&
0.5\cos(\alpha)(1-\cos(2\alpha))
-0.25\sin^2(2\alpha)
+0.5\cos(\alpha)(1-\cos(2\alpha))
\\
-\sin(\alpha)\cos(\alpha)
+0.5\sin(\alpha)(1+\cos(2\alpha))
&
0.5\sin(\alpha)\sin(2\alpha)
+0.25(1+\cos(2\alpha))^2
-0.5\sin(2\alpha)\sin(\alpha)
&
-0.5\sin(\alpha)(1-\cos(2\alpha))
-0.25(1+\cos(2\alpha))\sin(2\alpha)
+0.5\cos(\alpha)\sin(2\alpha)
\\
\sin^2(\alpha)
&
0.5\sin(\alpha)(1+\cos(2\alpha))
-\cos(\alpha)\sin(\alpha)
&
-0.5\sin(\alpha)\sin(2\alpha)
+\cos^2(\alpha)
\end{bmatrix},
$}
\endgroup
\]
where $\alpha = kX$. We find that $M^4_{\pi/2}=I$ for $\alpha=0$, which leads to $X=0$ for a non-zero value of $k$. This yields $\phi=\pi/2$ or $\theta=0,\pi$. If the dynamics starts from near these $\theta,\phi$, it returns to the initial position after $4$ periods. Any state lying on these orbits exhibits period-$4$ oscillations in the observables, as observed for the $4$-DTC phase in Fig.~\ref{fig_4dtc_am}. Therefore, unlike the $2$-DTC phase, the $4$-DTC phase appears only for particular values of $\theta$ and $\phi$ in the coherent state. In Fig.~\ref{fig_bifurcation}, we show two different initial conditions and their corresponding quantum and classical dynamical behaviors as a function of $k$. There exists a parameter regime of $k$ for $p=\pi/2$ in which $\theta$ exhibits four branches in the classical case, whereas $\langle J_z\rangle$ displays four branches in the quantum case. As this parametric regimes is crossed, the system exhibits chaotic behavior, as indicated in Fig.~\ref{fig_bifurcation}.

\section{QFI Curvature and Order Parameter for general states} 
\label{Appendix:B}
\begin{figure*}[htbp]
    \centering
    \includegraphics[width=0.975\textwidth, scale=1, trim=0 0 0 0, clip=true]{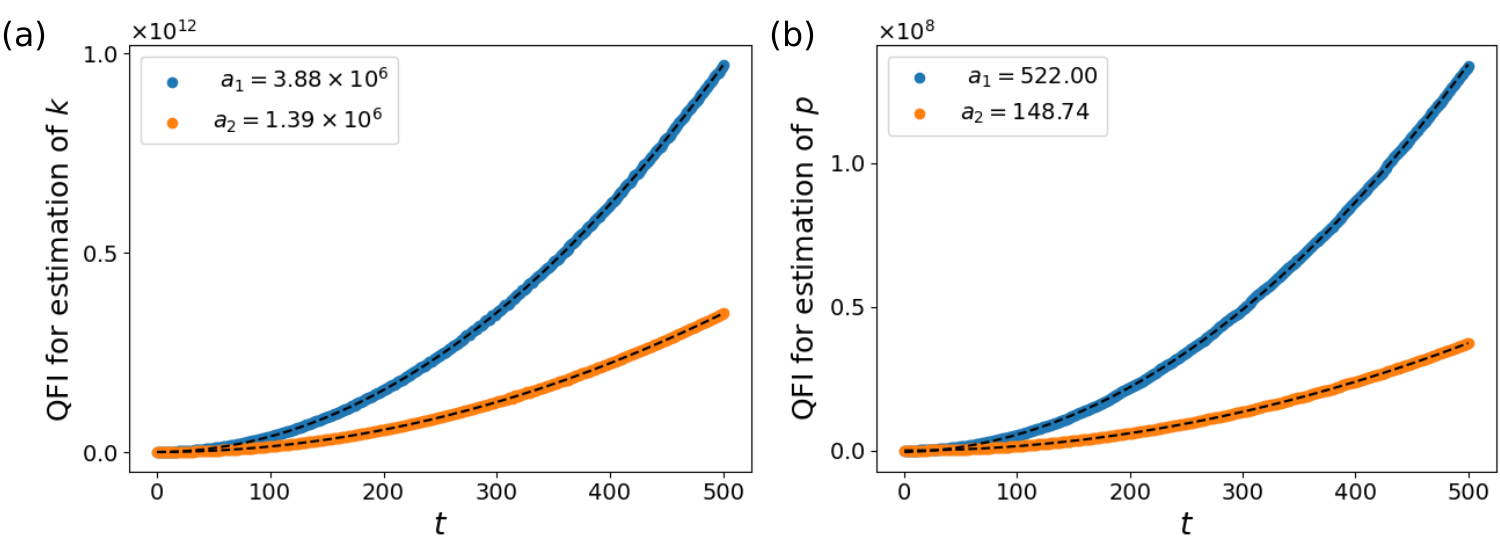}
    \caption{ Quadratic fit of QFI estimated for (a) $k$ for two points at edge of $2$-DTC with $p=3.74$ and $3.63$ (b) $p$ for two points at edge of $4$-DTC with $p=1.73$ and $1.65$ ; for $k=1.5$ and $j = 100$.}
    \label{fig_qfiqfit}
\end{figure*}

In this appendix, we demonstrate various dynamical phases by evaluating the order parameter ($O$) and standard deviation ($\Delta$), as defined in Sec.~\ref{DTC_DF}, for a generic coherent state to illustrate the general applicability of our definition. In the main text, however, we consider the specific state $|j,-j\rangle$. In addition, we calculate QFI curvatures of the corresponding phase plots. To investigate the growth rate of the QFI, we fit the data with a polynomial, although the long-time behavior of the QFI is known to be quadratic~\cite{Ullah2026}. The QFI can be written as
\begin{align*}
F_Q(t)=at^{2}+bt+c \ \ \  , \ \ \ \frac{d^{2}F}{dt^{2}}=2a=\kappa_{\lambda},
\end{align*}
where $a$, $b$ and $c$ are fitting parameters. As mentioned earlier, the coefficient of $t^2$ is referred to as the QFI curvature with respect to time. The quadratic behavior of QFI for estimating both $k$ and $p$ on the phase boundaries is shown in Fig.~\ref{fig_qfiqfit}.

\begin{figure*}[htbp]
    \centering
    \includegraphics[width=1.07\textwidth, scale=1, trim=0 222 0 0, clip=true]{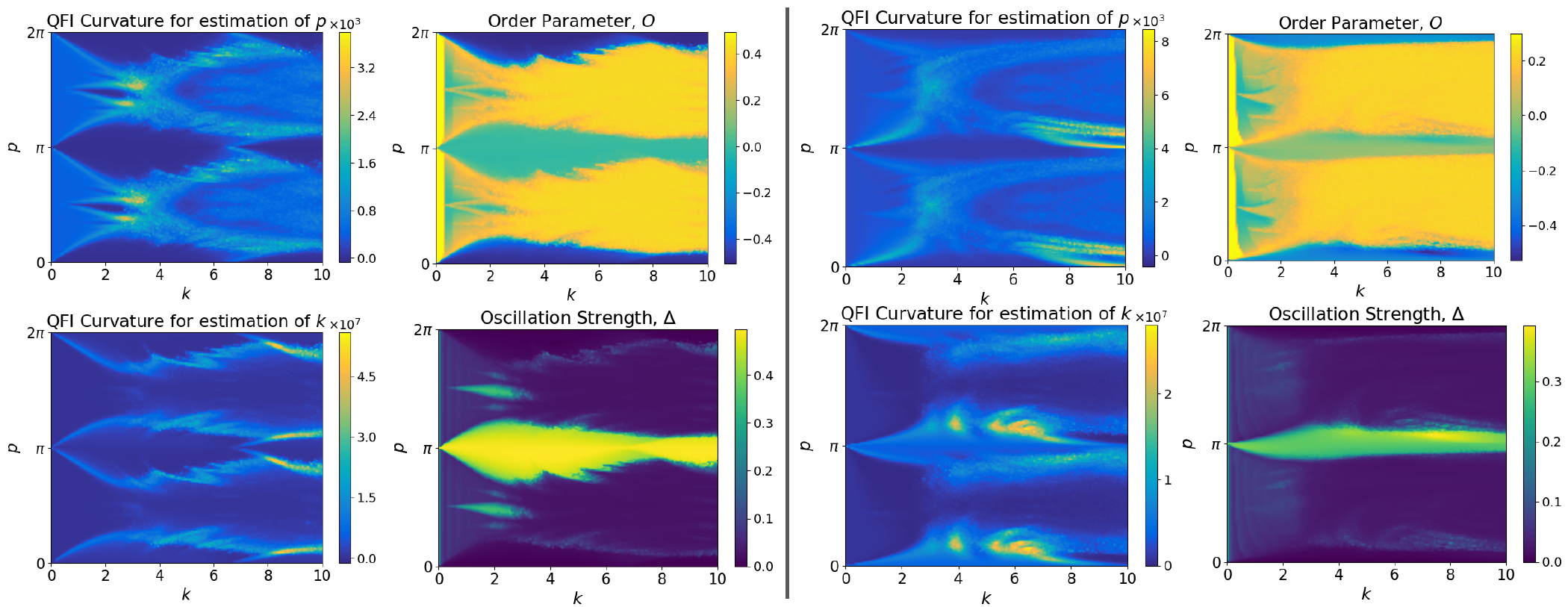}
    \caption{ Left set: Density plots for QFI curvatures, Order parameter ($O$) and Standard Deviation/Oscillation Strength ($\Delta$) corresponding the initial coherent state $|\theta=0.2,\phi=0.5\rangle$. Right set: Same as left one with the initial state $|\theta=2.20,\phi=0.77\rangle$. Here, $j = 100$.}
    \label{fig_correspondence}
\end{figure*}

In the main text, we present phase diagrams in the $p$–$k$ plane illustrating various dynamical phases through the behavior of the order parameter ($O$) and standard deviation ($\Delta$) for the particular state $|j,-j\rangle$. However, our definition of the order parameters is generic and can be applied to any initial state. Although the boundary regimes may change with the choice of the initial state, the overall physics remains unchanged. The same is illustrated in the left and right sets of plots in Fig.~\ref{fig_correspondence}. In the left set, the initial state is chosen such that the $4$-DTC phase exists, whereas in the right set, a different initial state is considered for which the $4$-DTC phase does not exist. For both cases, the order parameters and the QFI curvatures provide consistent results. This demonstrates the general applicability of the proposed order parameter and oscillation strength in identifying different dynamical phases of the system.

\end{document}